\documentclass[aps,prl,preprint,superscriptaddress,amsmath,amssymb,longbibliography]{revtex4-2}
\usepackage[latin1]{inputenc}
\usepackage{graphicx}% Include Fig. files
\usepackage{dcolumn}% Align table columns on decimal point
\usepackage{bm}% bold math
\usepackage{color}% colors in text
\usepackage{tabularx}
\newcolumntype{Y}{>{\centering}X}
\usepackage[cdot,textstyle,amssymb]{SIunits}

\usepackage{sistyle}
\SIunitspace{\cdot}
\SIunitdot{\cdot}
\SIdecimalsign{.}
\usepackage{ulem}
\usepackage[pdfborder={0 0 0}, colorlinks=true,citecolor=blue,linkcolor=black,urlcolor=blue]{hyperref}

 \definecolor{orange}{rgb}{1, 0.625, 0.0625}

\begin{document}

% Use the \preprint command to place your local institutional report
% number in the upper righthand corner of the title page in preprint mode.
% Multiple \preprint commands are allowed.
% Use the 'preprintnumbers' class option to override journal defaults
% to display numbers if necessary
%\preprint{}

%Title of paper
\title{Phononic-magnetic dichotomy of the thermal Hall effect in the Kitaev-Heisenberg candidate material Na$_2$Co$_2$TeO$_6$}

% repeat the \author .. \affiliation  etc. as needed
% \email, \thanks, \homepage, \altaffiliation all apply to the current
% author. Explanatory text should go in the []'s, actual e-mail
% address or url should go in the {}'s for \email and \homepage.
% Please use the appropriate macro foreach each type of information

% \affiliation command applies to all authors since the last
% \affiliation command. The \affiliation command should follow the
% other information
% \affiliation can be followed by \email, \homepage, \thanks as well.
%\author{}
%\email[]{Your e-mail address}
%\homepage[]{Your web page}
%\thanks{}
%\altaffiliation{}
%\affiliation{}

%Collaboration name if desired (requires use of superscriptaddress
%option in \documentclass). \noaffiliation is required (may also be
%used with the \author command).
%\collaboration can be followed by \email, \homepage, \thanks as well.
%\collaboration{}
%\noaffiliation

%%%%%%%%%%%%%%%%%%%%%%%%%%%%%%%%%%%%%%%%%%%%%%%%%%%%%%%%%%%%%%%%
													%Authors
%%%%%%%%%%%%%%%%%%%%%%%%%%%%%%%%%%%%%%%%%%%%%%%%%%%%%%%%%%%%%%%%

\author{Matthias Gillig}
\affiliation{Leibniz IFW Dresden, Helmholtzstr. 20, 01069 Dresden, Germany}

\author{Xiaochen Hong}
\affiliation{Leibniz IFW Dresden, Helmholtzstr. 20, 01069 Dresden, Germany}
\affiliation{Fakult\"at f\"ur Mathematik und Naturwissenschaften, Bergische Universit\"at Wuppertal, 42097 Wuppertal, Germany}

\author{Christoph Wellm}
\affiliation{Leibniz IFW Dresden, Helmholtzstr. 20, 01069 Dresden, Germany}

\author{Vladislav Kataev}
\affiliation{Leibniz IFW Dresden, Helmholtzstr. 20, 01069 Dresden, Germany}

\author{Weiliang Yao}
\affiliation{International Center for Quantum Materials, School of Physics, Peking University, 100871 Beijing, China}

\author{Yuan Li}
\affiliation{International Center for Quantum Materials, School of Physics, Peking University, 100871 Beijing, China}
\affiliation{Collaborative Innovation Center of Quantum Matter, 100871 Beijing, China}

\author{Bernd B\"uchner }
\affiliation{Leibniz IFW Dresden, Helmholtzstr. 20, 01069 Dresden, Germany}
\affiliation{Institute of Solid State Physics, TU Dresden, 01069 Dresden, Germany}
\affiliation{Center for Transport and Devices, TU Dresden, 01069 Dresden, Germany}

\author{Christian Hess}%\email{c.hess@uni-wuppertal.de}
\affiliation{Leibniz IFW Dresden, Helmholtzstr. 20, 01069 Dresden, Germany}
\affiliation{Center for Transport and Devices, TU Dresden, 01069 Dresden, Germany}
\affiliation{Fakult\"at f\"ur Mathematik und Naturwissenschaften, Bergische Universit\"at Wuppertal, 42097 Wuppertal, Germany}

%%%%%%%%%%%%%%%%%%%%%%%%%%%%%%%%%%%%%%%%%%%%%%%%%%%%%%%%%%%%%%%%

\date{\today}

% insert suggested keywords - APS authors don't need to do this
%\keywords{}

%\maketitle must follow title, authors, abstract, \pacs, and \keywords
\maketitle

% \section{Introduction}
\textbf{
Majorana fermions as emergent excitations of the Kitaev quantum spin liquid ground state constitute a promising concept in fault tolerant quantum computation. Experimentally, the recently reported topological half-quantized thermal Hall effect in the Kitaev material $\alpha$-RuCl$_3$ seems to confirm the Majorana nature of the material's magnetic excitations.
It has been argued, however, that the thermal Hall signal in $\alpha$-RuCl$_3$ rather stems from phonons or topological magnons than from Majorana fermions. Here we investigate the thermal Hall effect of the closely related Kitaev quantum material Na$_2$Co$_2$TeO$_6$, and we show that the thermal Hall signal emerges from at least two components, phonons and magnetic excitations. This dichotomy results from our discovery that the transversal heat conductivity $\kappa_{xy}$ carries clear signatures of the phononic $\kappa_{xx}$, but changes sign upon entering the low-temperature, magnetically ordered phase.  We systematically resolve the two components by considering the detailed temperature and field dependence of both $\kappa_{xy}$ and $\kappa_{xx}$. Our results demonstrate that uncovering a genuinely quantized magnetic thermal Hall effect in a Kitaev topological quantum spin liquid requires to disentangle phonon vs. magnetic contributions where the latter include potentially fractionalized excitations such as the expected Majorana fermions.}\\

\noindent The strongly frustrated Kitaev model describes spin-1/2 degrees of freedom on a honeycomb lattice with bond-dependent interactions \cite{Kitaev2006}. This exactly solvable spin model attracts strong attention in the community because it exhibits a quantum spin liquid (QSL) ground state with peculiar spin excitations which fractionalize into localized $Z_2$ gauge fluxes (also called visons) and itinerant Majorana fermions \cite{Kitaev2006,Baskaran2007,Knolle2014} which might become exploitable for quantum memories protected from decoherence \cite{Kitaev2006}.
Particular interest in this regard has been generated by the expectation that an external magnetic field renders the ground state a topological quantum spin liquid  with chiral Majorana edge currents and a field-induced bulk gap. These edge currents are expected to give rise to a thermal Hall signal in experiments \cite{Kitaev2006} with a half-quantized transversal thermal conductivity $\kappa_{xy}/T$.

For a few years, the compound $\alpha$-RuCl$_3$ has been considered a prime candidate material \cite{Plumb2014}, for which many experimental probes hint towards compatibility with the characteristics of Kitaev's model \cite{Motome2020}. Indeed, this compound exhibits a sizeable thermal Hall effect \cite{Kasahara2018,Kasahara2018a,Hentrich2019,Yamashita2020,Lefrancois2022,Bruin2022}. Interestingly, for certain ranges of temperature and magnetic field, a plateau has been reported in $\kappa_{xy}/T$ at a value which corresponds to exactly $1/2$ of the quantum of the two-dimensional thermal Hall conductance $K_{QH}/T=\frac{\pi^2k_B^2}{3h}$ \cite{Kasahara2018a,Yamashita2020,Bruin2022,Wolter2022}, a result which could be recognized as the key fingerprint of the topologically protected Majorana edge currents. However, there are concerns about the uniqueness of this finding, and topological magnons \cite{Czajka2023} as well as chiral phonons \cite{Hentrich2019,Lefrancois2022} have been suggested as alternative origins of the thermal Hall effect in $\alpha$-RuCl$_3$.

\begin{figure}[t]
\centering
\includegraphics[width=\columnwidth]{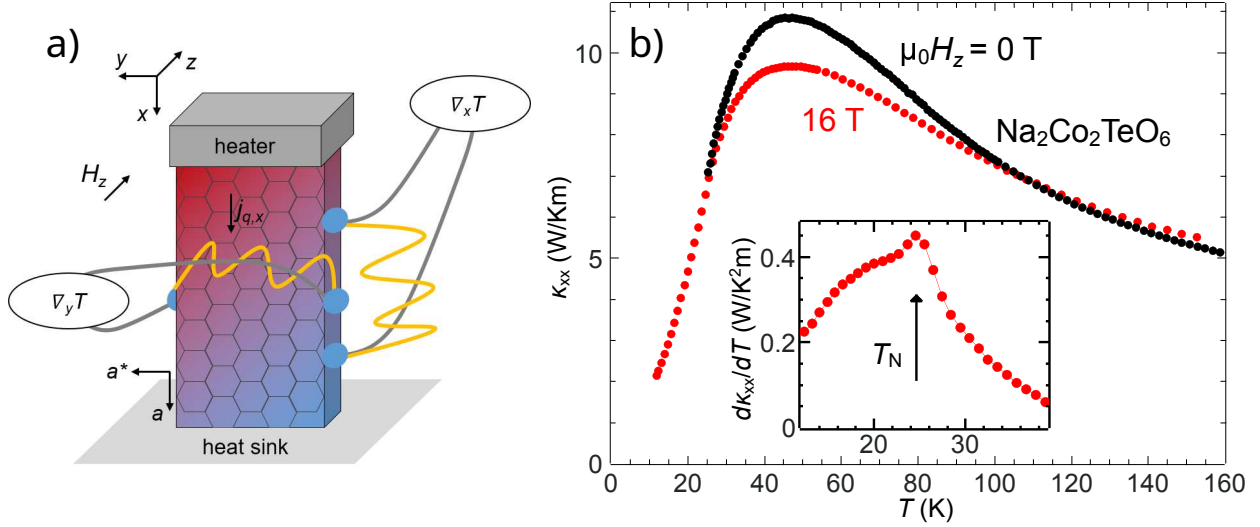}
\caption{Experimental setup scheme and longitudinal heat conductivity data of Na$_2$Co$_2$TeO$_6$. a) Experimental setup indicating the orientations of the thermal current density $j_{q,x}$ and the longitudinal temperature gradient $\nabla_x T$ parallel to the $a$-axis of the compound and the magnetic field $H_z$ along the $c$-axis (perpendicular to the planes). The transverse temperature gradient $\nabla_y T$ is measured perpendicular to the thermal curent and the magnetic field. b) Temperature dependence of the longitudinal heat conductivity $\kappa_{xx}$ of Na$_2$Co$_2$TeO$_6$ at zero field and at a magnetic field $\mu_0H_z = 16$~T. The inset shows the temperature derivative $d\kappa_{xx}/dT$ of the longitudinal thermal conductivity at 16~T.}
\label{fig:kappavst}
\end{figure}

Recently, Na$_2$Co$_2$TeO$_6$ has been proposed as a novel possible materialization of Kitaev physics in a honeycomb lattice \cite{Liu2018,Yao2020}, originating from the expectation of a bond dependent Ising-like interaction between the pseudospin-1/2 of the $d^7$ Co$^{2+}$ ions in high-spin $t^5_{2g}e^2_g$ configuration \cite{Liu2018,Sano2018}. The material orders antiferromagnetically below $T_N\approx27$~K, and a zigzag \cite{Lefrancois2016} as well as a triple-\textbf{q} \cite{Chen2021May, Yao2022Nov, Kruger2022Nov} ground state have been proposed following neutron scattering data. The magnetically ordered ground state implies the significance of Heisenberg and/or off-diagonal interactions additional to the expected primary Kitaev coupling \cite{Chaloupka2010,Chaloupka2013}. An in-plane magnetic field has been found to suppress the magnetic order at $\mu_0 H> \mu_0 H_c\approx10$~T, in favour of strong low-energy spin fluctuations at fields near $H_c$ which are compatible with a field-induced QSL \cite{Hong2021}. Altogether, these features bear an astonishing similarity with the magnetic phase diagram of $\alpha$-RuCl$_3$ \cite{Hong2021}. 
Here, we investigate the longitudinal and the transversal thermal conductivity (the thermal Hall effect), i.e. $\kappa_{xx}$ and  $\kappa_{xy}$, with the thermal current in the honeycomb plane parallel to the $a$ axis and a magnetic field perpendicular to the plane (see Fig.~\ref{fig:kappavst}a and methods).\\

%%%%%%%%%%%%%%%%%%%%%%%%%%%%%%%%%%%%%%%%%%%%%%%%%%%%%

%%%%%kappa_xx vs T
\noindent\textbf{Temperature dependence of $\kappa_{xx}$ and $\kappa_{xy}$}\\
% We first investigate the impact of a magnetic field along the $c$ axis on the longitudinal thermal conductivity $\kappa_{xx}$. 
Fig.~\ref{fig:kappavst}b shows a comparison of  $\kappa_{xx}$ at zero magnetic field, which is very similar to our previous findings \cite{Hong2021}, and at $\mu_0 H= 16$~T with ${\bf H}\| c$. Apparently, apart from a slight reduction of about 11\%, the overall curve shape is preserved even at this relatively large field.
Only a weak change is also observed concerning the onset temperature of magnetic ordering which we infer from the cusp in $d\kappa_{xx}/dT$ shown in the inset of Fig.~\ref{fig:kappavst}b as $T_N\approx25$~K, which is only slightly lower than in zero field \cite{Hong2021,Yao2020}. Note that an in-plane magnetic field has been reported to have a much stronger impact: $\kappa_{xx}$ becomes strongly enhanced while the magnetic order is suppressed for $\mu_0H\gtrsim 10$~T \cite{Hong2021}.

\begin{figure}[t]
\centering
\includegraphics[width=0.8\columnwidth]{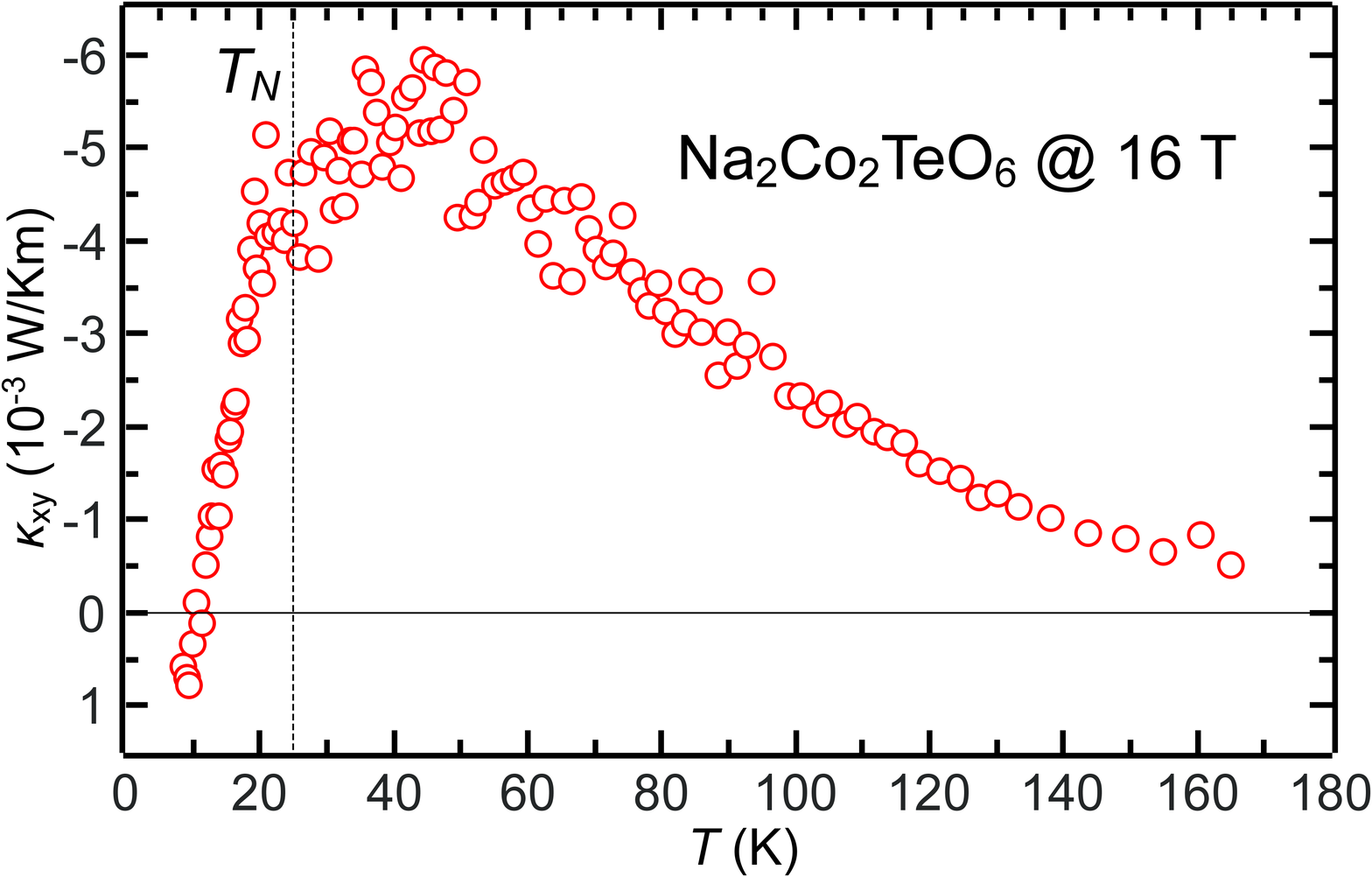}
\caption{
Temperature dependence of the transverse heat conductivity $\kappa_{xy}$ of Na$_2$Co$_2$TeO$_6$ with the heat current $j_{q,x}$ along the $a$-axis at zero field and at a magnetic field $\mu_0 H_z = 16$~T along the $c$-axis. }
\label{fig:kappa-xy-vs-T}
\end{figure}

%%%%\kappa_xy vs T
%\textbf{Transversal thermal conductivity $\kappa_{xy}$ as a function of temperature}\\
Fig.~\ref{fig:kappa-xy-vs-T} presents our findings for the transversal heat conductivity $\kappa_{xy}$ at 16~T as a function of temperature. It reaches a maximum value of $\kappa_{xy} \approx -6 \cdot 10^{-3}~\mathrm{W/Km}$ which is about three times smaller than the predicted half-integer quantized value for Na$_2$Co$_2$TeO$_6$ at 40~K ($\kappa_{xy} \approx 19.9 \cdot 10^{-3}~\mathrm{W/Km}$). At first glance one can recognize strong similarities between $\kappa_{xx}$ and $-\kappa_{xy}$ (as plotted in the figure). Both curves rapidly rise with increasing temperature $T$, exhibit a broad peak at about 40-50~K, and then gradually decay to lower values with further rising $T$. However, different from the longitudinal thermal conductivity which only can have positive values, $\kappa_{xy}$ changes sign at about 12~K from positive values at $T\lesssim12$~K to negative at higher $T$. 

%%%%%%%%%%%%%%%%%%%%%%%%%%%%%%%%%%%%%%%%%%%%%%%%%%%%%

\begin{figure}[t]
\centering
\includegraphics[width=0.8\columnwidth]{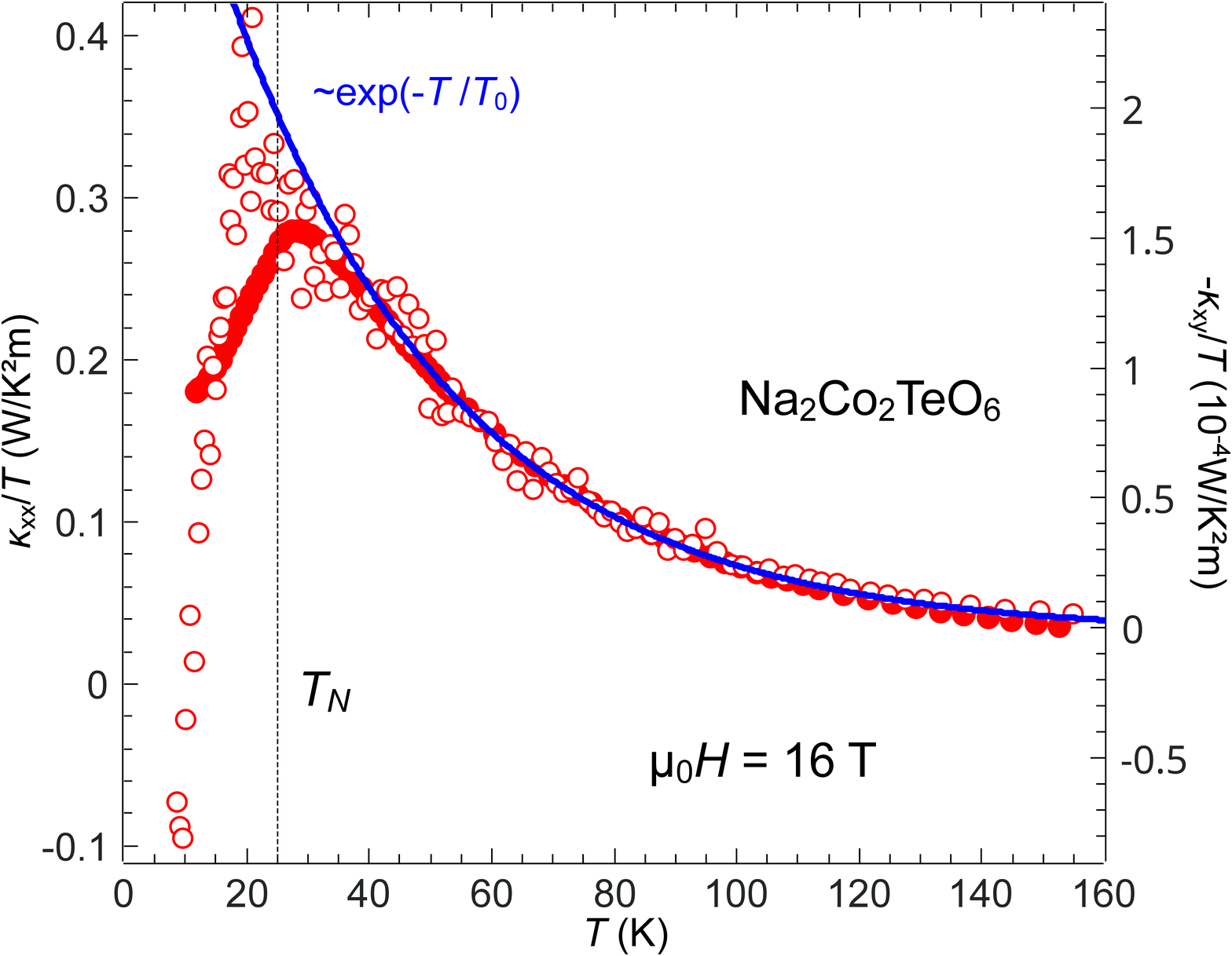}
\caption{Comparison of $\kappa_{xy}/T$ (open symbols, right scale) and $\kappa_{xx}/T$ (full symbols, left scale) of Na$_2$Co$_2$TeO$_6$ at $\mu_0H=16$~T directed along the $c$-axis. The solid line shows a fit according to $\kappa_{xy}/T=A\exp(-T/T_0) + B$ for $T>35$~K with the parameters $A=-3.93\cdot 10^{-4}~\mathrm{W/K^2m}$, $B=2.06\cdot10^{6}~\mathrm{W/K^2m}$, and $T_0=36.8$~K.}
\label{fig:kappa-xx-xy-vst}
\end{figure}

%%%%%\kappa_xx/T vs kappa\kappa_xy/T
%\textbf{Comparison $\kappa_{xx}/T$ and $\kappa_{xy}/T$}\\
In order to disentangle possible origins of the observed thermal Hall effect (magnetic or phononic), we further investigate in Fig.~\ref{fig:kappa-xx-xy-vst} the temperature dependence of $-\kappa_{xy}/T$. Remarkably, in this representation,  a strong cusp-like anomaly is present close to the magnetic ordering temperature $T_N$, which separates a region $T\lesssim T_N$ with a very steep temperature dependence including the sign change from the high temperature region $T\gtrsim T_N$ where $\kappa_{xy}/T$ decays moderately with rising temperature. 

We first focus on the latter, paramagnetic regime. Here the transversal heat conductivity very well follows an exponential scaling $\kappa_{xy}/T\sim\exp(-T/T_0)$ which recently has been proposed as a generic feature for the temperature dependence of the thermal Hall effect of charge neutral excitations \cite{Yang2020}. According to the modeling the observation of this scaling suggests a nontrivial topology of the heat carrying quasiparticles with a finite and essentially temperature independent effective Berry curvature density. Unfortunately, the exponential scaling is expected to universally hold for all types of quasiparticles and thus it does not allow to draw clear-cut conclusions about the type of the excitation causing the thermal Hall effect.
However, a direct comparison of $-\kappa_{xy}/T$ with $\kappa_{xx}/T$ (see Fig.~\ref{fig:kappa-xx-xy-vst}) strikingly yields a perfect match of both curves for $T>T_N$, i.e., in the whole paramagnetic regime.
The only difference is here a temperature independent offset since $\kappa_{xy}/T\rightarrow 0$ at high temperature whereas $\kappa_{xx}/T$ approaches a finite value. 
Thus, in view of the almost perfect matching temperature dependencies of $\kappa_{xx}/T$ and $\kappa_{xy}/T$ and the fact that the heat carrying quasiparticles contributing to the longitudinal $\kappa_{xx}$ are phonons (see Ref.~\cite{Hong2021} and discussion further below), it is tempting to interpret this observation as evidence for a phononic origin of the thermal Hall for $T\gtrsim T_N$.
%%%%%%%%%%%%%%%%%%%%%%%%%%%%%%%%%%%%%%%%%%%%%%%%%%%%%
%%%%%%%%%%%%%%%%%%%%%%%%%%%%%%%%%%%%%%%%%%%%%%%%%%%%%
Indeed, this conclusion seems to be corroborated by findings for $\alpha$-RuCl$_3$ where the similarities between $\kappa_{xx}/T$ and $\kappa_{xy}/T$ have been interpreted as significant evidence for a phononic origin of the thermal Hall effect \cite{Lefrancois2022}. Furthermore, the recent demonstrations of phonon thermal Hall effect in several magnetic insulators supports the notion that similar temperature dependencies of $\kappa_{xy}/T$ and a phononic $\kappa_{xx}/T$ are a general fingerprint of phononic thermal Hall effect \cite{Grissonnanche2020,Boulanger2020,Chen2022Aug}. \\

%%%%%%%T<TN
%\textbf{Origin of the sign change at low T}\\
However, the low-$T$ behavior of $\kappa_{xy}$ in the magnetically ordered regime is difficult to be explained by a sole phononic contribution to $\kappa_{xy}$.
As can be inferred clearly from the data in Fig.~\ref{fig:kappa-xx-xy-vst}, $\kappa_{xx}/T$ and $\kappa_{xy}/T$ exhibit distinctly different temperature dependencies at $T\lesssim T_N$. While upon cooling $\kappa_{xx}/T$ is only moderately reduced by about 30\% at the lowest temperature measured, $-\kappa_{xy}/T$ decreases much steeper and changes sign at about 12~K. 
This sign change is a crucial information which implies either of two scenarios for the origin of the thermal Hall effect.

It is thinkable, on one hand, that $\kappa_{xy}$ indeed is purely phononic as inferred above. If this were the case the mechanism causing the off-diagonal thermal response has to fundamentally change upon cooling through $T_N$ and further upon possible reorientations of the magnetic order \cite{Hong2021} at even lower temperatures. Given the rather complicated and barely understood magnetic phase diagram of Na$_2$Co$_2$TeO$_6$ \cite{Yao2020,Hong2021,Xiao2021,Zhang2022Dec}, one cannot straightforwardly exclude this possibility without further information. One should note, however, that up to present all reported examples of phonon thermal Hall effect do not exhibit a sign change \cite{Boulanger2020,Chen2022Aug,Grissonnanche2019,Grissonnanche2020,Hirokane2019,Ideue2017,Li2020,Sugii2017,Li2023,Ataei2023Feb}. Hence this scenario seems rather unlikely.

On the other hand, if one assumes that the sign of the phonon thermal Hall effect is always \textit{negative}, the only possibility is that a second \textit{positive }contribution to $\kappa_{xy}$ rapidly gains importance, causing the sign change of $\kappa_{xy}$ in the magnetically ordered phase. In this case, this second contribution must be due to transport by magnetic degrees of freedom. Among these, within this paper we do not distinguish further since magnetic fluctuations remain strong even at low temperature \cite{Hong2021} and hence both magnons as well as possible fractionalized excitations such as Majorana fermions and visons a priori can not be excluded. \\

%%%%%%%%%%%%%%%%%%%%%%%%%%%%%%%%%%%%%%%%%%%%%%%%%%%%%
\begin{figure}[thb]
\centering
\includegraphics[width=\columnwidth]{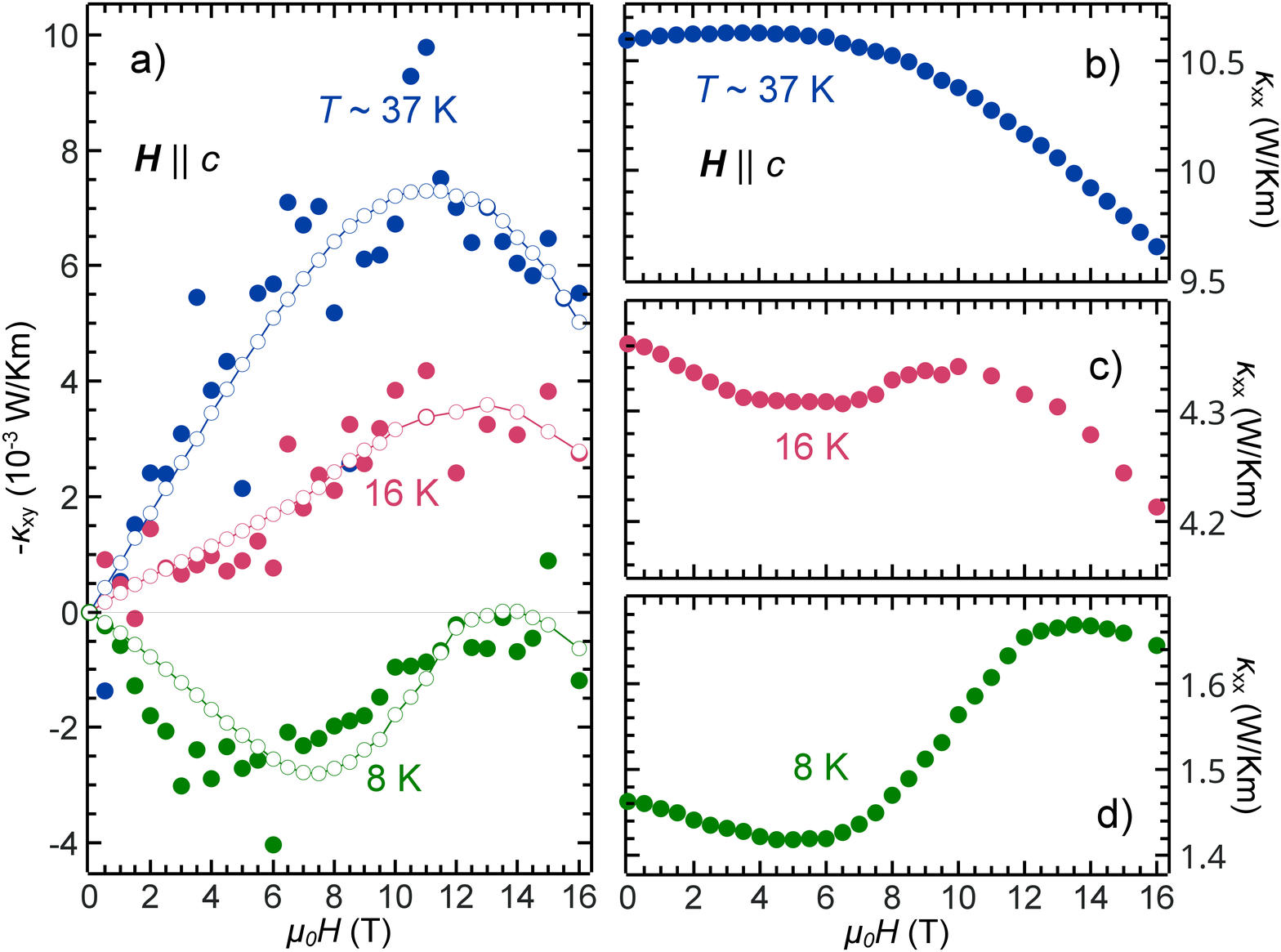}
\caption{Field dependence of the longitudinal and transversal thermal conductivities of Na$_2$Co$_2$TeO$_6$ for selected temperatures.
a) Field dependence of the transverse thermal conductivity $\kappa_{xy}$. Open symbols represent a fit according to Eq. \ref{eq:kappa_xy} using $\kappa_{xx}$-data. b)-d) Field dependence of the longitudinal thermal conductivity $\kappa_{xx}$.}
\label{fig:kappa-xx-xy-vsb}
\end{figure}
%%%%%%%%%%%%%%%%%%%%%%%%%%%%%%%%%%%%%%%%%%%%%%%%%%%%%

% \noindent\textbf{Field dependence of $\kappa_{xx}$ and $\kappa_{xy}$}\\

\noindent\textbf{Magnetic field dependence of $\kappa_{xx}$ and $\kappa_{xy}$}\\
The latter scenario of a two-component thermal Hall conductivity is strongly supported by the isothermal magnetic field
dependence of $\kappa_{xy}$ as presented in Fig.~\ref{fig:kappa-xx-xy-vsb}a) at selected temperatures 8~K, 16~K, and 37~K. As can be clearly inferred from the data, $\kappa_{xy}(H)$ at all these temperatures exhibits a pronounced and mostly non-monotonic field dependence. This is quite unusual since a non-monotonic field dependence is in strong contrast to other findings on various materials including $\alpha$-RuCl$_3$ where a quasi linear field dependence of $\kappa_{xy}(H)$ has been observed \cite{Strohm2005,Sugii2017,Grissonnanche2019,Li2020,Kasahara2018,Hentrich2019,Li2023}.

The origin of the unusual $\kappa_{xy}(H)$ can straightforwardly be connected with the field dependence of the isothermally measured $\kappa_{xx}(H)$  as shown in Fig.~\ref{fig:kappa-xx-xy-vsb}b)-d)  as we will discuss in detail further below. 
% Here, we first describe and discuss here $\kappa_{xx}(B)$ for various temperatures.
$\kappa_{xx}(H)$ possesses two features with opposite trends in temperature. One is a pronounced minimum at about 4-8~T which is strongest at 8~K, very weak at 16~K and absent at 37~K. The other is a decrease at higher field with increasing field.
% with increasing field at $\mu_0 H\gtrsim10$~T. 
% This decrease is weak at 8~K, where it is discernible only for $\mu_0H>14$~T
At 8~K the decrease is weak and discernible only for $\mu_0H>14$~T, but with raising the temperature it becomes increasingly important. At 16 K it begins at around 10~T, and at 37~K it dominates the data for $\mu_0H\gtrsim4$~T. Clearly, the growing importance is consistent with the data shown in Fig.~\ref{fig:kappavst}b. 

These $\kappa_{xx}(H)$ data are characteristic of a phonon heat conductivity which is subject to strong phonon-spin scattering \cite{Jeon2016,Hentrich2018,Hentrich2020, Gillig2021Dec}.
In fact, $\kappa_{xx}$ has already been shown to be governed by phonons with strong phonon-spin scattering for magnetic fields applied in plane \cite{Hong2021}. In the situation which we consider here, i.e. with a magnetic field perpendicular to the magnetic planes, for the $(H, T)$-dependence of $\kappa_{xx}$ one should consider two qualitative phonon-spin scattering mechanisms: In the magnetically ordered state $T\leq T_N$ one can expect that the phonon spin scattering becomes particularly important (i) near field-driven spin reorientation transitions \cite{Gillig2021Dec} and (ii) "resonant" scattering of the phonons off collective excitations such as magnons \cite{Hentrich2018, Hentrich2020,Hong2021}.

Concerning the former, an initial version of the magnetic $(H, T)$-phase diagram for $H||c$ \cite{Zhang2022Dec} reveals a practically temperature independent magnetic transition at about 8~T for $T\lesssim16$~K. Associated spin fluctuations are therefore a plausible cause for the dip observed in $\kappa_{xx}(H)$ at 8 and 16~K.

On the other hand clear-cut information about low-energy magnon excitations in zero field is available from recent inelastic neutron scattering (INS) data \cite{Yao2022}. The lowest lying magnon mode has been reported in the energy range 1-3 meV with minima at the
$\Gamma$ and the $M$ points. It is thus clear, if one considers a typical $\Theta_D$ of a few hundred K, that phonon-magnon scattering involving the acoustic phonon modes must be significant already at zero field. In order to obtain further insight in the field and temperature dependence of
the magnon mode and $\kappa_{xx}(H)$, we investigated the magnetic field induced energy shift of this magnon mode at the $\Gamma$-point by means of high-field electron spin resonance (ESR, see Supplementary Information). At $T = 3$~K, the data yield an energy gap at the $\Gamma$-point of about 219~GHz (0.91~meV) at zero field, in accordance with the zero field
neutron result, and a Zeeman shift to about 950~GHz (about 3.9 meV) at 14~T (corresponding to a $g$ factor of 3.91).

If one assumes a rigid magnon band shift in magnetic field the maximum magnon energy at 16~T amounts to about 6.8~meV corresponding to about 80~K.
Standard considerations for phonon heat conductivity \cite{Callaway1961}, namely 

\begin{equation} 
	\kappa_{\text{ph}}\propto T^3\int_0^{\Theta_D/T} \frac{x^4\exp(x)}{(\exp(x)-1)^2}\tau_c dx, \label{eq:kappa_ph}
\end{equation}

with $\tau_c$ a total phononic relaxation time imply phonons with an energy of about $4k_BT$ to dominate the heat conductivity. 
Thus, within a simple kinematic picture for the phonon-magnon scattering, for temperatures smaller than some characteristic $T_\mathrm{peak}< (80~\mathrm{K}/4)\lesssim 20$~K one would naturally expect a minimum in $\kappa_{xx}(H)$ because in this $T$-range the magnetic field should drive the 'intersection' of the phonon and magnon bands through the peak of the integrand in Eq.~\ref{eq:kappa_ph}. At higher temperatures, the applied magnetic field is only sufficient to drive the band intersection \textit{towards} the integrand peak without actually reaching it and therefore the data should exhibit only a suppression with increasing field. Both the minimum at low $T$ and the increasing high-field suppression is qualitatively observed, superimposing on the effect of the phase transition described above. 

The above considerations should be understood within some margin of error: The actual $T_\mathrm{peak}$ and further details of $\kappa_{xx}(H, T)$ certainly depend on the true $\Theta_D$, the magnon density of states and the effects of temperature and magnetic field on the magnon dispersion. Concerning the latter, it is worth pointing out that our ESR data imply that at $T = 6$-8~K the magnon mode starts to increasingly soften and broaden with increasing the temperature consistent with zero magnetic field results of INS \cite{Yao2022}. The broadening gets particularly strong at $T > 20$~K, still below $T_N$, possibly due to enhancement of the spin fluctuations by approaching the phase transition. It is plausible that for $T>T_N$ a qualitatively similar scenario applies for paramagnon excitations.

After having established the phonon-only nature of $\kappa_{xx}(H)$ and the connection of its anomalous field dependence to phonon-magnon scattering we turn now to $\kappa_{xy}(H)$ and directly compare the data for the longitudinal and transverse heat transport channels. Very clearly, at 8~K both curves for  $\kappa_{xx}(H)$ and $-\kappa_{xy}(H)$ possess a very similar relative field dependence, i.e. a minimum at 4-6~T and a saturation at $\mu_0 H\gtrsim12$~T. At the higher temperatures, 16~K and 37~K the connection between $\kappa_{xx}(H)$ and $-\kappa_{xy}(H)$  at first glance seems not so clear as seen for 8~K. However, a closer inspection reveals that for all data the derivative $\partial\kappa_{xy}(H)/\partial H$ is proportional to $\kappa_{xx}(H)$ modulo a constant offset. In order to demonstrate this revealing connection we plot in Fig.~\ref{fig:kappa-xx-xy-vsb}a) a fit according to 

\begin{equation}
 \kappa_{xy}(H)=a_1\cdot \kappa_{xx}(H)\cdot \mu_0 H+a_2\cdot \mu_0 H \label{eq:kappa_xy}
\end{equation}
with $a_1$ and $a_2$ as fitting constants. As can be seen in the figure, this empirical ansatz describes the data very well, apart from a slight low-field deviation at 8~K, yielding parameters $a_1 < 0$ and $a_2 > 0$ (see Supplementary Information). 

Eq.~\ref{eq:kappa_xy} clearly suggests that $\kappa_{xy}$ is composed of two additive components with opposite signs. 
Due to its weighting with the phononic $\kappa_{xx}$, the first term can be assigned to a phononic thermal Hall effect $\kappa_{xy,\mathrm{ph}}$.
On the other hand, the second term is strictly linear in magnetic field, which corresponds to the usual observation for $\kappa_{xy}(H)$  for the vast majority of compounds, see above. This component must be of a different origin than $\kappa_{xy,\mathrm{ph}}$. The most straightforward conclusion is thus that this positive component is magnetic in nature ($\kappa_{xy,\mathrm{mag}}$). Thus, the measured total thermal Hall effect should be understood as the sum of the phononic and magnetic contributions to the thermal Hall effect, i.e. $\kappa_{xy}=\kappa_{xy,\mathrm{ph}}+\kappa_{xy,\mathrm{mag}}$. The observed sign change in the measured $\kappa_{xy}$ indicates a change in the dominance of one component with respect to the other.

%%%%%%%%%%%%%%%%%%%%%%%%%%%%%%%%%%%%%%%%%%%%%%%%%%%%%
\begin{figure}[thb]
	\centering
	\includegraphics[width=0.8\columnwidth]{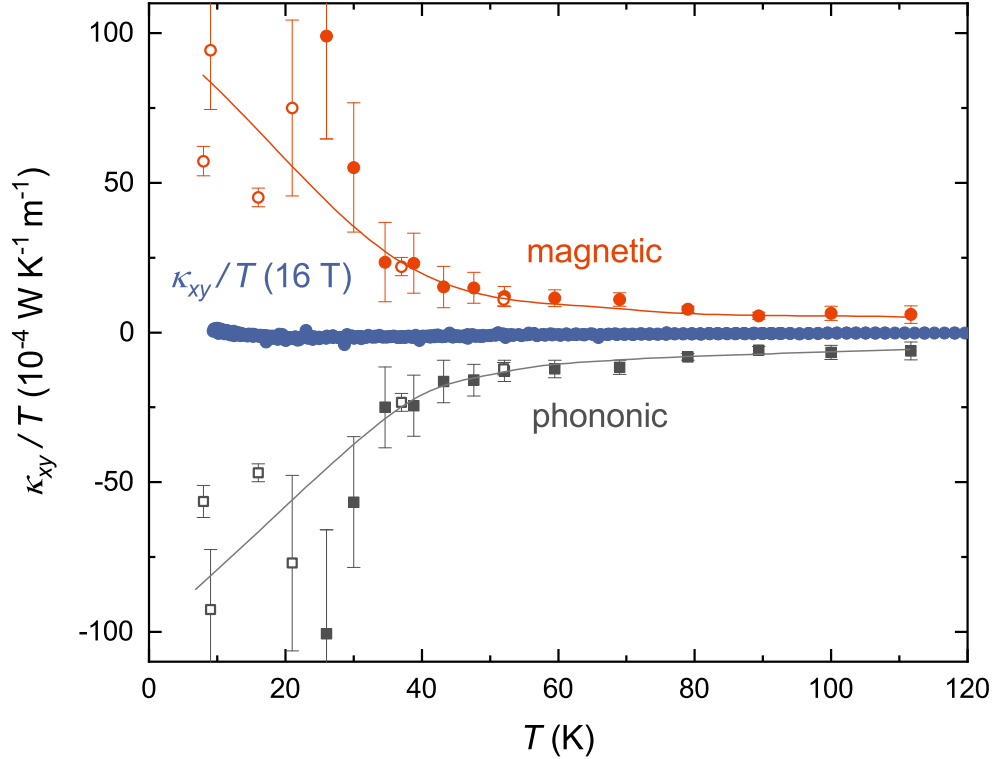}
	\caption{ Decomposition of $\kappa_{xy}/T$ of Na$_2$Co$_2$TeO$_6$ at 16~T in a negative phononic and a positive magnetic contribution according to fitting results of Eq. \ref{eq:kappa_xy} compared to the experimental data (blue). Open symbols refer to fits of the field dependence whereas filled symbols represent fits to data of temperature dependent measurements. Solid lines are guides to the eye.}
	\label{fig:kappa-xy-phon_mag}
\end{figure}
%%%%%%%%%%%%%%%%%%%%%%%%%%%%%%%%%%%%%%%%%%%%%%%%%%%%%

It is instructive to extract the separate temperature dependences of $\kappa_{xy,\mathrm{ph}}$ and $\kappa_{xy,\mathrm{mag}}$, which straightforwardly can be achieved for selected temperatures from the fits shown in Fig.~\ref{fig:kappa-xx-xy-vsb}. In addition, we use two  temperature dependent data sets for both $\kappa_{xx}$ and $\kappa_{xy}$ measured at $\mu_0H=6$~T and 16~T (see Supplementary Information), and Eq.~\ref{eq:kappa_xy} for extracting $\kappa_{xy,\mathrm{ph}}$ and $\kappa_{xy,\mathrm{mag}}$ at 16~T. The complete data set is shown in Fig.~\ref{fig:kappa-xy-phon_mag}  in comparison with the total measured $\kappa_{xy}$ as a function of temperature. This comparison clearly reveals that for $T>T_N$ the absolute value of $\kappa_{xy,\mathrm{ph}}$ always is somewhat larger than that of $\kappa_{xy,\mathrm{mag}}$, yielding the overall negative sign. However, upon the onset at $T<T_N$, this fine balance changes and $\kappa_{xy,\mathrm{mag}}$ exceeds $|\kappa_{xy,\mathrm{ph}}|$,  leading to the sign change of $\kappa_{xy}$ at low temperature.

% \section{Discussion and summary}
Our finding of a two component thermal Hall effect, i.e. with a phononic and a magnetic contribution should be placed into the context of recent theoretical and experimental findings. First of all, our conclusion of a sizeable $\kappa_{xy,\mathrm{ph}}$ is well compatible with several recent theoretical works for magnetic systems \cite{Sheng2006,Qin2012,Ye2020,Ye2021} where phonon-spin scattering has been identified as one important source for generating a finite phononic transverse thermal conductivity. Indeed, as demonstrated above, phonon-spin scattering is the primary scattering process in the phononic $\kappa_{xx}$ at all temperatures considered. 
On the other hand, a magnetic thermal Hall effect has been shown to be expected for chiral magnons in Kitaev magnets \cite{McClarty2018}, in addition to the expectation of a thermal Hall effect from Majorana fermion edge currents \cite{Kitaev2006,Nasu2017}. The growing importance of $\kappa_{xy,\mathrm{mag}}$ observed  at $T<T_N$ in our data seems indeed compatible with an additional effect due to chiral magnons.
We point out that a microscopic mechanism as well as a quantitative and qualitative prediction of the phononic and magnetic thermal Hall effect in Na$_2$Co$_2$TeO$_6$ and more generally in Kitaev systems still needs to be worked out. On this general scheme, our results also provide fresh input for understanding the controversially discussed thermal Hall effect in $\alpha$-RuCl$_3$. The experimental significance of a phononic thermal Hall effect as shown in our data for Na$_2$Co$_2$TeO$_6$ imply that a sizeable phononic thermal Hall effect should be expected in  $\alpha$-RuCl$_3$, too, because both systems possess a very similar phonon-spin scattering phenomenology \cite{Hentrich2018,Hong2021}. Therefore, our results support the recent notion of a phononic thermal Hall effect in $\alpha$-RuCl$_3$ \cite{Lefrancois2022} and call for a reinvestigation of the intriguing findings for a low-temperature plateau \cite{Kasahara2018a,Bruin2022}.
More specifically, the key for uncovering a genuinely quantized magnetic thermal Hall effect should be to disentangle phonon vs. magnetic contributions to the $\kappa_{xy}$, and to look for quantized behavior only in the magnetic part, rather than in the total $\kappa_{xy}$, which does not deserve to be quantized with the phonon contribution.

%%%%%%%%%%%%%%%%%%%%%%%%%%%%%%%%%%%%%%%%%%%%%%%%%%%%%%%%%%%%%%%%\'{e}\c{c}

\section{Methods}
High-quality Na$_2$Co$_2$TeO$_6$ single crystals were grown by a modified flux method \cite{Yao2020,Xiao2019}. A regular bar-shaped sample of $5.05 \times 1.03 \times 0.10$~mm$^3$ was cut from an as grown crystal with the sample edges parallel to the $a\text{-}$, $a^*\text{-}$, and $c$-axes, respectively. The cut sample was mounted in a home-built probe, employing a 6-points measurement geometry, see Fig.~\ref{fig:kappavst}a. In our configuration, a thermal current density $j_{q,x}$ was generated parallel to the $a$-axis (i.e. the zigzag direction of the honeycomb lattice) using a chip heater. The thereby produced longitudinal thermal gradient $\nabla_x T$ was measured using a field-calibrated differential Au/Fe-Chromel thermocouple. The transverse temperature gradient $\nabla_y T$ along the $a^*$-direction which arises upon applying a magnetic field parallel to the $c$-axis, i.e. perpendicular to the honeycomb layers, was measured by a second thermocouple of the same type. 

The longitudinal $\nabla_x T$ and the transversal $\nabla_y T$ were measured simultaneously. The small size of the transverse signal required the application of a large heat current to achieve a reasonable signal-to-noise ratio, resulting in temperature differences in the order of $\nabla_x T/T_0$ = 10~\% and $\nabla_y T/T_0$ = 0.01~\% compared to the thermal bath temperature $T_0$. At fixed temperature the heater current was varied to prove a linear behavior in $\nabla_y T$.
To consider significant heating the sample temperature was determined by extrapolating $T_0$ to the position of the transverse thermocouple in the center of the sample.
To eliminate longitudinal contributions to the transverse signal due to a possible misalignment of the thermocouple contacts, measurements were performed under both field polarities and longitudinal components have been eliminated by antisymmetrization of $\nabla_y T$.

High-field high-frequency electron spin resonance (HF-ESR) experiments were carried out in a frequency range $250-950$\,GHz using a home-made multifrequency spectrometer. For the generation and detection of the microwave radiation a vector network analyzer (PNA-X from Keysight Technologies), as well as a combination of a modular Amplifier/Multiplier Chain  (AMC from Virginia Diodes, Inc.) and  a hot electron InSb bolometer (QMC Instruments) were employed. A single crystal of Na$_2$Co$_2$TeO$_6$  was mounted into a probe head operational in the transmission mode that was put in a $^4$He variable temperature inset of a  superconducting magnet system (Oxford Instruments) producing fields up to 16\,T (see also Ref.~\cite{Alfonsov2021}).

% Create the reference section using BibTeX:
% \bibliographystyle{apsrev4-1}
%\bibliography{/home/chris/Dokumente/Literatur/NCTO/ncto,/home/chris/Dokumente/Literatur/RuCl/rucl,/home/chris/Dokumente/Literatur/Thermal_Hall/thermalhall,/home/chris/Dokumente/Literatur/kappaphon/kappa_ph}
% \begingroup
% \renewcommand{\section}[2]{References}%
\noindent\bibliography{ncto,rucl,thermalhall,kappa_ph,ESR}
% \endgroup
%merlin.mbs apsrev4-1.bst 2010-07-25 4.21a (PWD, AO, DPC) hacked
%Control: key (0)
%Control: author (8) initials jnrlst
%Control: editor formatted (1) identically to author
%Control: production of article title (-1) disabled
%Control: page (0) single
%Control: year (1) truncated
%Control: production of eprint (0) enabled

\newpage
% \vspace{1cm}

\noindent\textbf{Acknowledgments}\\
This work has been supported by the Deutsche Forschungsgemeinschaft through SFB 1143 (project-id 247310070), through the projects HE3439/13 and through grant No. KA 1694/12-1. The work at Peking University was supported by the National Basic Research Program of China (Grants No. 2021YFA1401900 and No. 2018YFA0305602).\\

\noindent\textbf{Author Contributions}\\
M.G. and X.H. performed the thermal transport measurements, C.W. and V.K. performed the ESR measurements, W.Y. and Y.L grew the samples, B.B. and C.H. designed the research, M.G. and C.H. analysed the data. All
authors discussed the data analysis and interpretation. M.G. and C.H. wrote the paper.\\

\noindent\textbf{Competing interests}\\
The authors declare no competing interests.\\

\noindent\textbf{Data availability}\\
The data supporting the findings of this study are available from the corresponding author
upon reasonable request. \\

\noindent\textbf{Author information}\\
Correspondence and request for materials should be addressed C.H. (c.hess@uni-wuppertal.de).

\newpage
\section{Supplementary information}

\renewcommand\thefigure{S\arabic{figure}} 
\setcounter{figure}{0}  
\renewcommand\thetable{S\arabic{table}} 
\subsection{Electron spin resonance}

\begin{figure}[htb]
	\centering
	\includegraphics[clip,width=0.9\columnwidth]{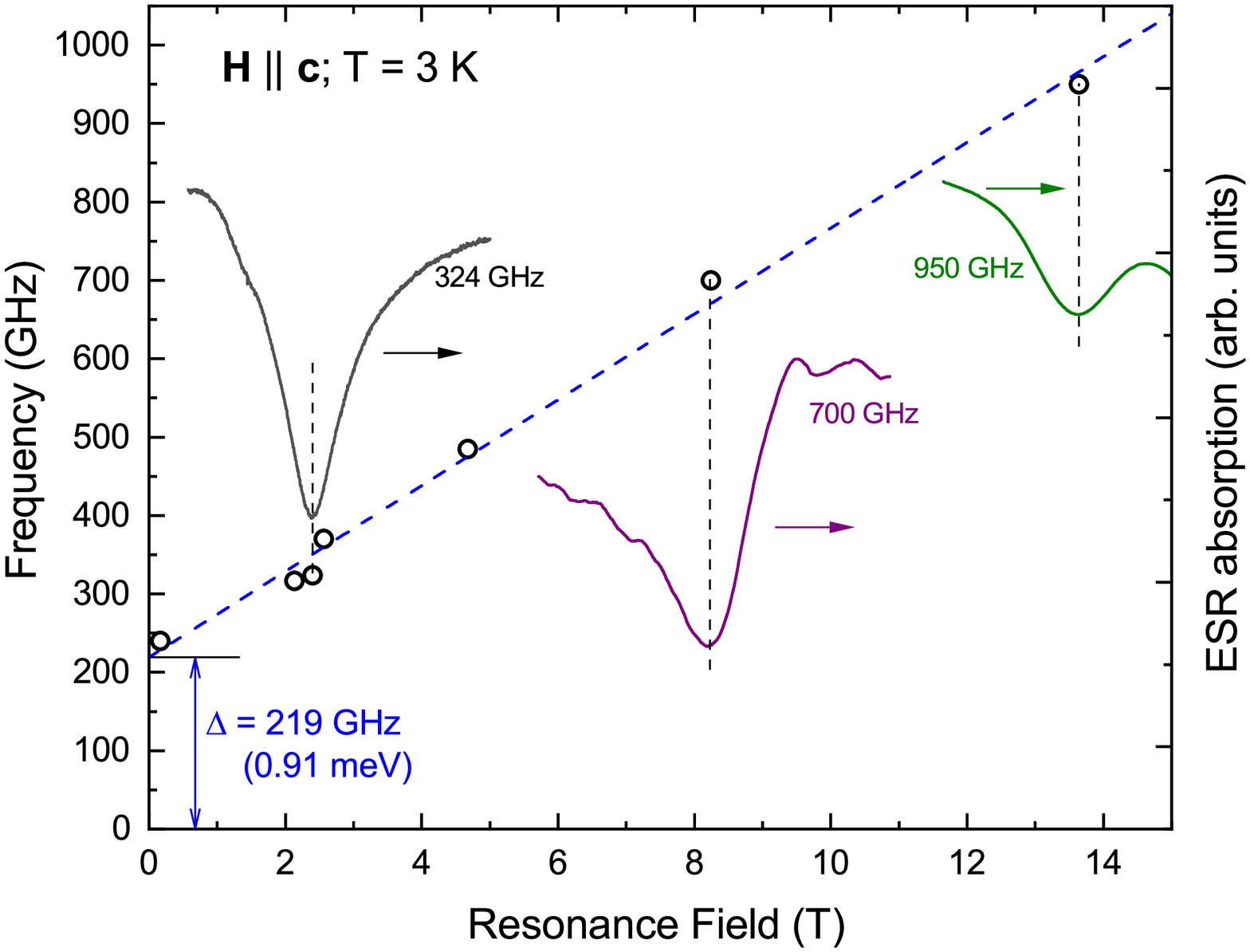}
	\caption{Frequency $\nu$ {\it versus} resonance field $H_{\rm res}$ dependence of the ESR signal of Na$_2$Co$_2$TeO$_6$ for the ${\bf H}\parallel c$~axis field geometry at $T = 3$\,K (left vertical scale) together with representative HF-ESR spectra at selected frequencies (right vertical scale). 
		The dashed line depicts the linear dependence $\nu = \Delta + h^{-1}g_{\rm c}\mu_0\mu_{\rm B}H$ with the $g$~factor $g_{\rm c} = 3.91$ and the zero field magnon gap $\Delta = 219$\,GHz (0.91\,meV). }
	\label{fig:ESR_freq}
\end{figure}

The HF-ESR spectrum of Na$_2$Co$_2$TeO$_6$ in the magnetically ordered state for the magnetic field applied parallel to the crystal $c$ axis consists of a single absorption line corresponding to a uniform $k = 0$ magnon excitation of an antiferromagnetically ordered spin lattice. The resonance field $H_{\rm res }$ of this line scales linearly with the excitation frequency $\nu$ and follows the relation $\nu = \Delta + h^{-1}g_{\rm c}\mu_0\mu_{\rm B}H$ (Fig.~\ref{fig:ESR_freq}). From the slope of this dependence the $c$~axis component of the $g$~factor tensor can be determined amounting to $g_{\rm c} = 3.91$, the value typical for a Co$^{2+}$ ion in a distorted octahedral ligand coordination (see., e.g., Refs.~\cite{Wellm2021,Iakovleva2022}). Importantly, the data reveals a zero field excitation gap $\Delta = 219$\,GHz (0.91\,meV) which corresponds well with the magnitude of the magnon gap at the $\Gamma$ point in the 2D Brillouin  zone found in a recent inelastic neutron scattering work in Ref.~\cite{Yao2022}. 

\begin{figure}[htb]
	\centering
	\includegraphics[clip,width=0.6\columnwidth]{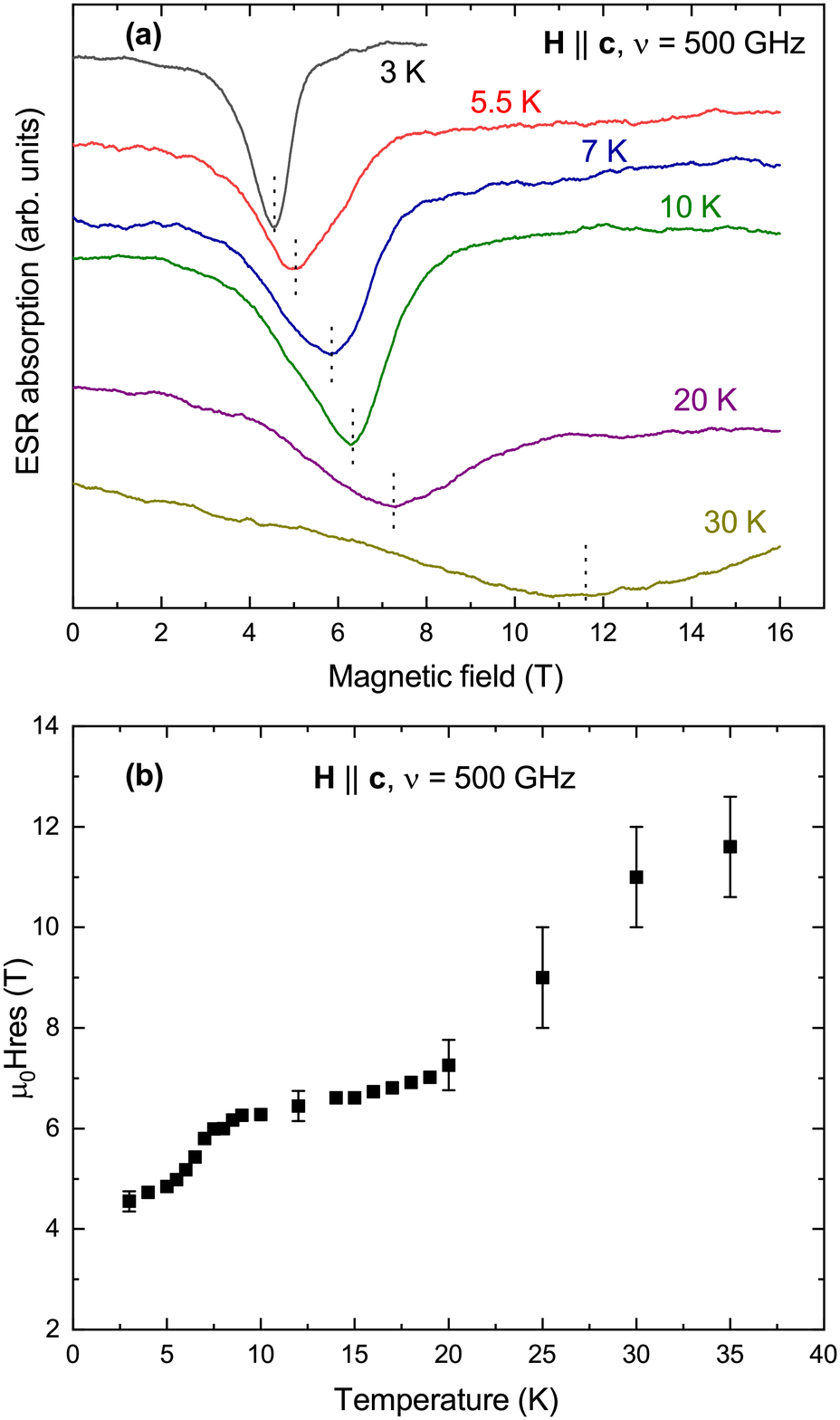}
	\caption{Temperature dependence of the ESR spectrum at $\nu =500$\,GHz (a) for ${\bf H}\parallel c$~axis (a) and of the resonance field $H_{\rm res}$ of the ESR line (b).}
	\label{fig:ESR_T}
\end{figure}

As can be seen in Fig.~\ref{fig:ESR_T}, for a given fixed excitation frequency, this $k = 0$ magnon mode shifts with increasing temperature to higher magnetic fields, obviously due to a softening (decreasing) of the magnon gap $\Delta$. By approaching the N\'eel ordering temperature $T_{\rm N} \approx 27$\,K this mode substantially broadens and becomes practically unobservable at $T > T_{\rm N}$, possibly due to the thermal population of the excited multiplet states of the Co$^{2+}$ ion \cite{Wellm2021}.

\subsection{Analysis of $\kappa_{xy}$}

\begin{figure}[htb]
	\centering
	\includegraphics[width=0.7\columnwidth]{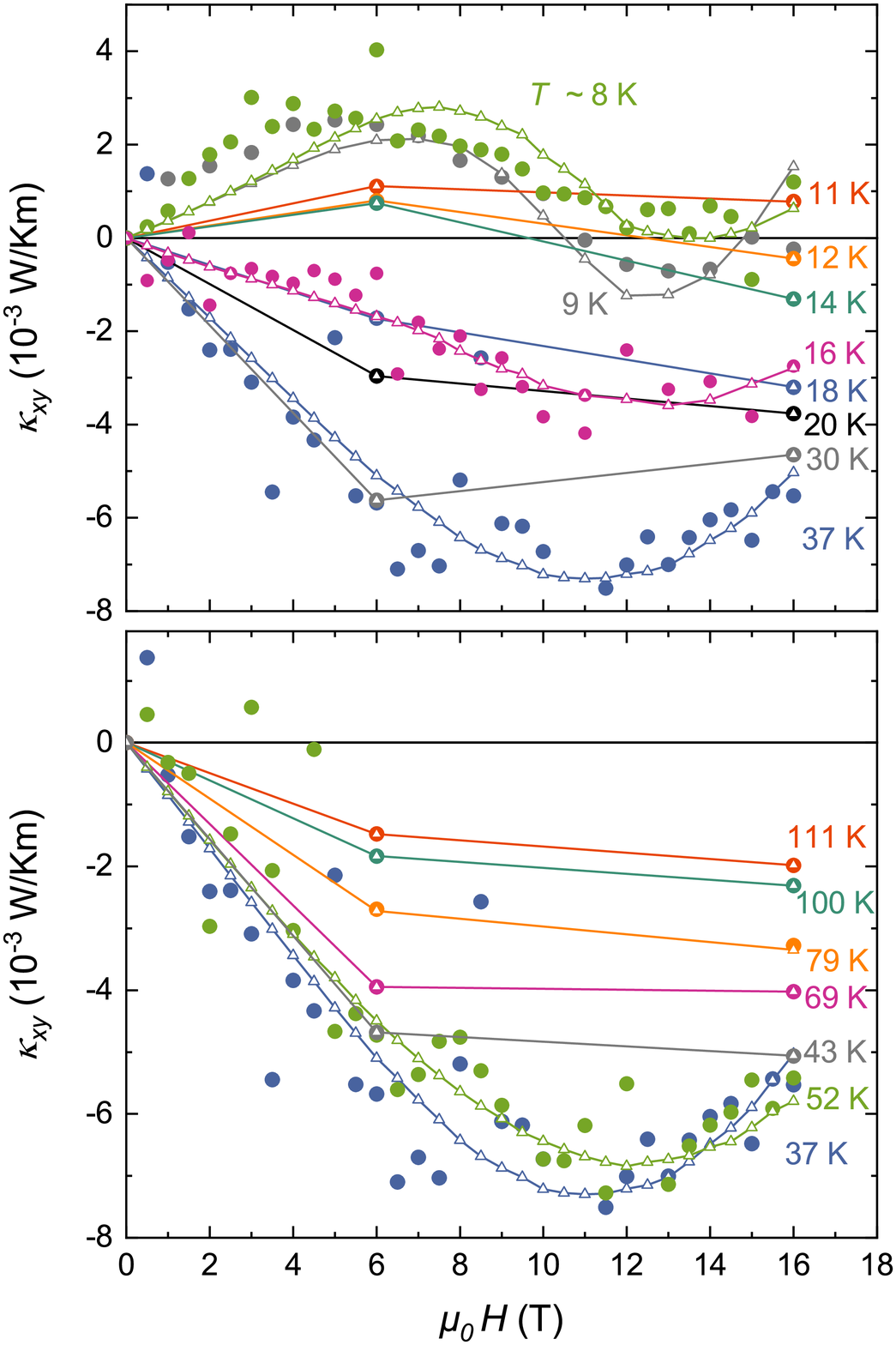}
	\caption{Field dependence of the thermal Hall effect of Na$_2$Co$_2$TeO$_6$ for selected temperatures where filled symbols represent experimental data and open triangles stand for fits according to Eq. \ref{eq:kappa_xy} using $\kappa_{xx}$-data. Solid lines are guides to the eye.}
	\label{fig:kappa-xy-vsH-fits}
\end{figure}

The thermal Hall effect data is analyzed according to the empirical equation \ref{eq:kappa_xy}.
The explicit field dependence of $\kappa_{xy}$ is fitted at 8, 9, 16, 21, 37 and 52~K (open symbols in Fig.~\ref{fig:kappa-xy-vsH-fits}). Error bars in Fig.~\ref{fig:kappa-xy-phon_mag} represent the difference between weighting the data at high and low field, above and below 8~T, respectively.

\begin{figure}[htb]
	\centering
	\includegraphics[width=0.6\columnwidth]{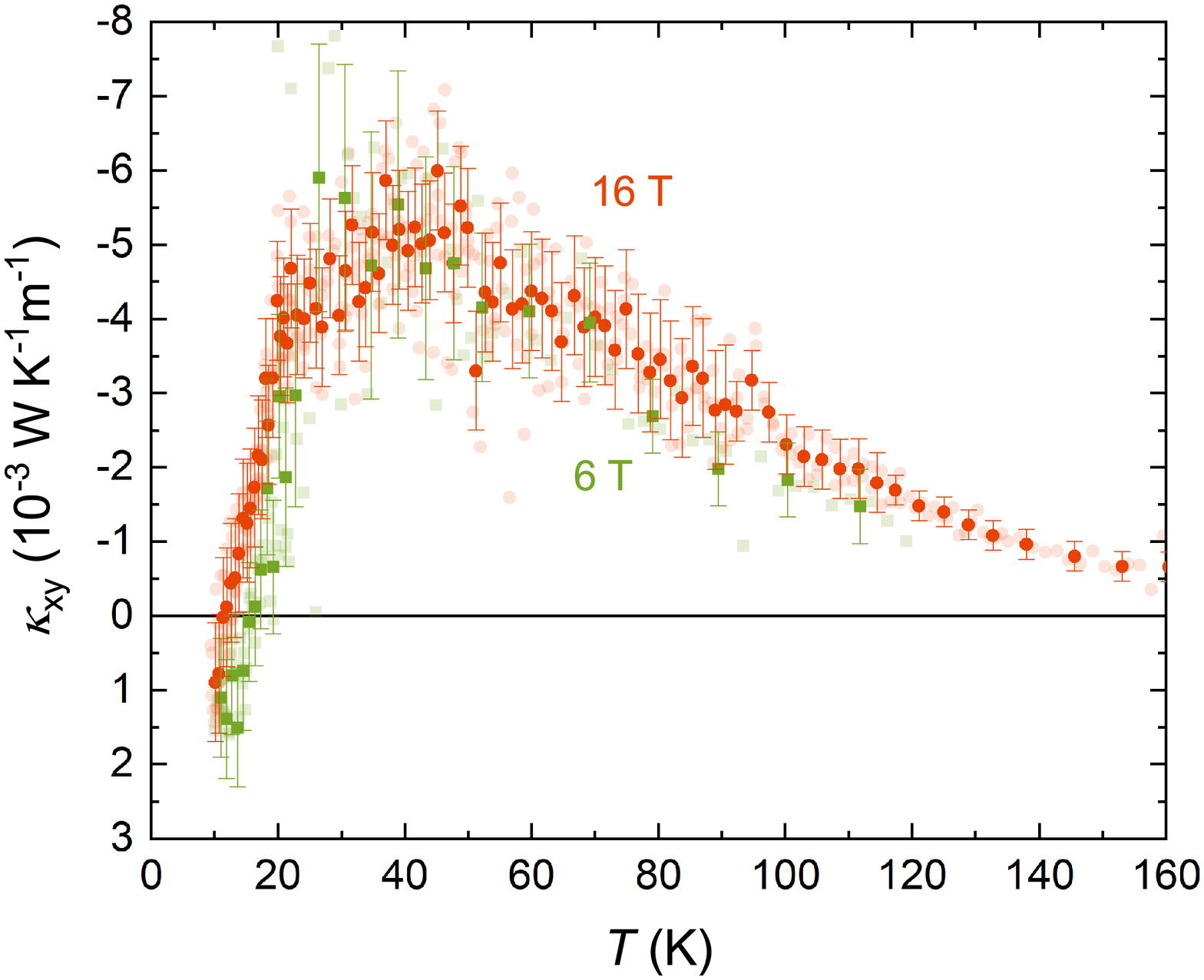}
	\caption{Temperature dependence of the thermal Hall effect of Na$_2$Co$_2$TeO$_6$ for magnetic field applied along the $c$-axis. Bold symbols represent the four-point average of the data (transparent symbols) and error bars account the spread before averaging.}
	\label{fig:kappa-xy-vsT}
\end{figure}

\begin{figure}[bth]
	\centering
	\includegraphics[width=0.6\columnwidth]{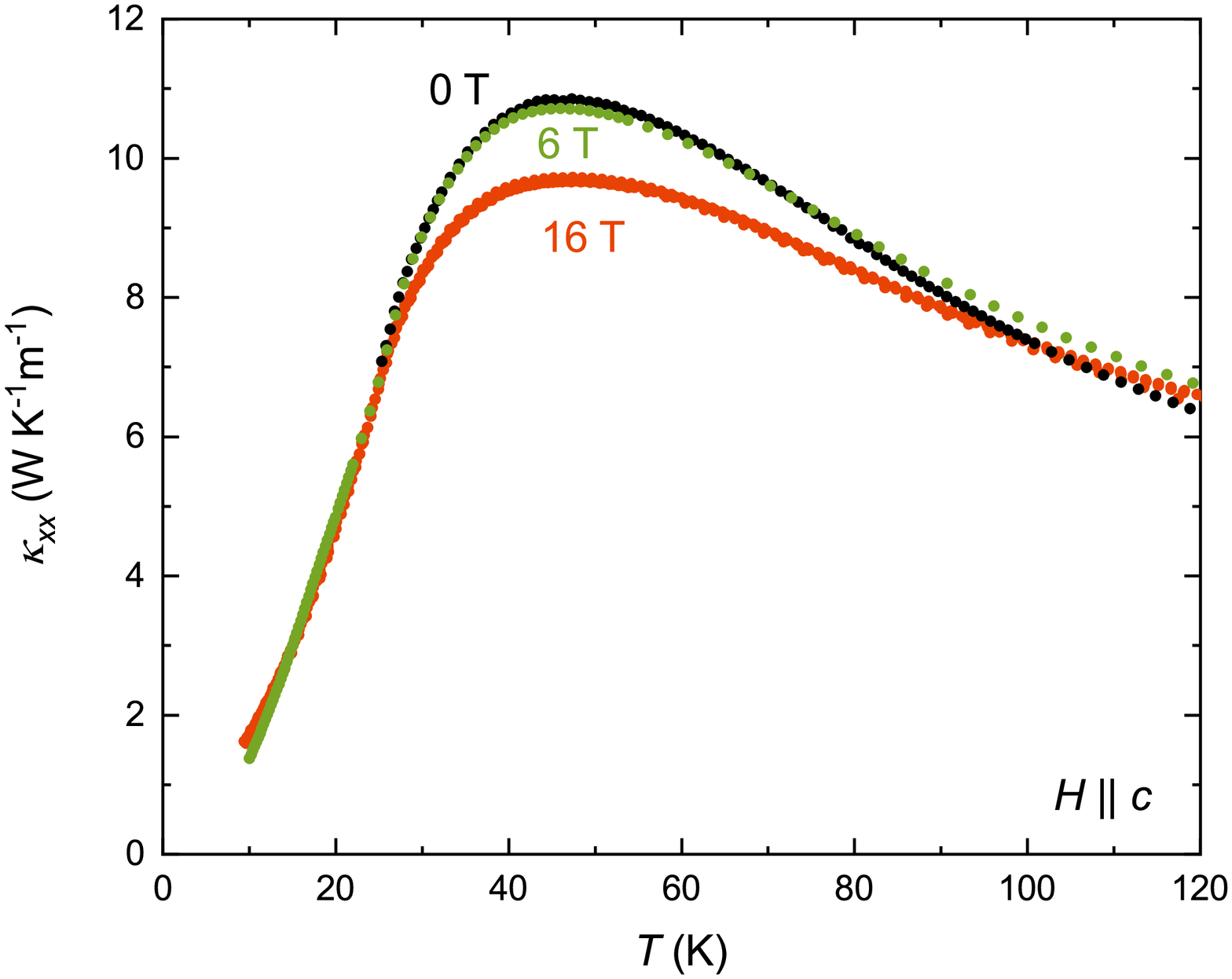}
	\caption{Temperature dependence of the thermal conductivity of Na$_2$Co$_2$TeO$_6$ for magnetic field applied along the $c$-axis.}
	\label{fig:kappa-xx-vsT}
\end{figure}

For temperatures above $T_N$, constant temperature cuts of $\kappa_{xy}(T)$-data (FIG. \ref{fig:kappa-xy-vsT}) and $\kappa_{xx}$-data (FIG. \ref{fig:kappa-xx-vsT}) for 0, 6 and 16~T were fit separately. Error bars take into account the spread of $\kappa_{xy}$-data before averaging by a highest and lowest deviation from the mean value.

%%%%%%%%%%%%%%%%%%%%%%%%%%%%%%%%%%%%%%%%%%%%%%%%%%%%%
%%%%%%%%%%%%%%%%%%%%%%%%%%%%%%%%%%%%%%%%%%%%%%%%%%%%%

%%%%%%%%%%%%%%%%%%%%%%%%%%%%%%%%%%%%%%%%%%%%%%%%%%%%%
%%%%%%%%%%%%%%%%%%%%%%%%%%%%%%%%%%%%%%%%%%%%%%%%%%%%%

%%%%%%%%%%%%%%%%%%%%%%%%%%%%%%%%%%%%%%%%%%%%%%%%%%%%%
%%%%%%%%%%%%%%%%%%%%%%%%%%%%%%%%%%%%%%%%%%%%%%%%%%%%%

\begin{table}[htb]
	\begin{tabular}{|ccc|}
		\hline
		$T $ & $a_1  $ & $a_2   $ \\
		(K) & ($10^{-3}$ T$^{-1}$) & ( $10^{-3}$ W(TKm)$^{-1}$) \\
		\hline
		8   & -1.716 &  2.862 \\
		9   & -2.314 &  5.304 \\
		16  & -0.482 &  1.739 \\
		21  & -1.751 &  9.845 \\
		26  & -2.170 & 16.096 \\
		30  & -1.253 & 10.336 \\
		35  & -0.592 &  5.087 \\
		37  & -0.560 &  5.089 \\
		39  & -0.625 &  5.613 \\
		43  & -0.458 &  4.106 \\
		48  & -0.489 &  4.444 \\
		52  & -0.412 &  3.581 \\
		52  & -0.434 &  3.909 \\
		60  & -0.482 &  4.263 \\
		69  & -0.558 &  4.750 \\
		79  & -0.474 &  3.813 \\
		89  & -0.417 &  3.126 \\
		100 & -0.565 &  4.017 \\
		112 & -0.625 &  4.179 \\
		\hline
		
	\end{tabular}
	\caption{Fit parameters from parameterization of $\kappa_{xy}$ according to Eq. \ref{eq:kappa_xy}}
\end{table}

\newpage

%%%%%%%%%%%%%%%%%%%%%%%%%%%%%%%%%%%%%%%%%%%%%%%%%%%%%%%%%%%%%%%%
%Bibliography
%%%%%%%%%%%%%%%%%%%%%%%%%%%%%%%%%%%%%%%%%%%%%%%%%%%%%%%%%%%%%%%%

% Create the reference section using BibTeX:
% \bibliographystyle{apsrev4-1}
%\bibliography{/home/chris/Dokumente/Literatur/NCTO/ncto,/home/chris/Dokumente/Literatur/RuCl/rucl,/home/chris/Dokumente/Literatur/Thermal_Hall/thermalhall,/home/chris/Dokumente/Literatur/kappaphon/kappa_ph}
% \bibliography{ncto,rucl,thermalhall,kappa_ph,ESR}
%merlin.mbs apsrev4-1.bst 2010-07-25 4.21a (PWD, AO, DPC) hacked
%Control: key (0)
%Control: author (8) initials jnrlst
%Control: editor formatted (1) identically to author
%Control: production of article title (-1) disabled
%Control: page (0) single
%Control: year (1) truncated
%Control: production of eprint (0) enabled

\end{document}